\documentclass[journal]{IEEEtran}
\ifCLASSINFOpdf
\else
\fi
\usepackage{graphicx,subcaption}
\usepackage{amssymb}
\usepackage{amsmath}
\usepackage{lineno}
\usepackage{algorithm}
\usepackage{algpseudocode}
\usepackage[hyphens]{url}
\usepackage{color}
\usepackage{xcolor}
\usepackage{mathtools}
\usepackage{lineno}
\usepackage{array}
\usepackage{pdflscape}
\usepackage{afterpage}
\usepackage{multirow}
\usepackage{booktabs}
\usepackage{balance}
\usepackage[pdfencoding=auto]{hyperref}
\usepackage{makecell}
\usepackage{siunitx}
\usepackage[version=4]{mhchem}
\usepackage{longtable}
\usepackage[export]{adjustbox}
\usepackage{enumitem}
\usepackage{cite}
\expandafter\def\expandafter\UrlBreaks\expandafter{\UrlBreaks
  \do\a\do\b\do\c\do\d\do\e\do\f\do\g\do\h\do\i\do\j%
  \do\k\do\l\do\m\do\n\do\o\do\p\do\q\do\r\do\s\do\t%
  \do\u\do\v\do\w\do\x\do\y\do\z\do\A\do\B\do\C\do\D%
  \do\E\do\F\do\G\do\H\do\I\do\J\do\K\do\L\do\M\do\N%
  \do\O\do\P\do\Q\do\R\do\S\do\T\do\U\do\V\do\W\do\X%
  \do\Y\do\Z}

\newcolumntype{L}[1]{>{\raggedright\let\newline\\\arraybackslash\hspace{0pt}}m{#1}}
\newcolumntype{C}[1]{>{\centering\let\newline\\\arraybackslash\hspace{0pt}}m{#1}}
\newcolumntype{R}[1]{>{\raggedleft\let\newline\\\arraybackslash\hspace{0pt}}m{#1}}

\usepackage{rotating}
\usepackage{amsmath}
\usepackage[nodayofweek]{datetime}
\newdateformat{mytoday}{\monthname[\THEMONTH] \twodigit{\THEDAY}, \THEYEAR}
\usepackage{mathtools}
\usepackage{array}
\usepackage{tabulary}
\newcolumntype{K}[1]{>{\centering\arraybackslash}p{#1}}

\ifCLASSINFOpdf
\else
\fi
\begin{document}
%
\title{Automatically Assessing Quality of Online Health Articles}
%
%
%

\author{Fariha~Afsana,~\IEEEmembership{Student Member,~IEEE,}  Muhammad~Ashad~Kabir,~\IEEEmembership{ Member,~IEEE,} Naeemul~Hassan, Manoranjan~Paul,~\IEEEmembership{Senior Member,~IEEE,}  


\thanks{F. Afsana, M. A. Kabir and M. Paul are with the School of Computing and Mathematics, Charles Sturt University, Australia}

\thanks{N. Hassan is with the Philip Merrill College of Journalism, University of Maryland, USA}

}
%
%

\markboth{IEEE Journal of Biomedical and Health Informatics, August~2019}%
{Afsana \MakeLowercase{\textit{et al.}}: Quality assessment of Health Articles}
%



\maketitle

\begin{abstract}
Information ecosystem today is overwhelmed by unprecedented quantity of data on versatile topics are with varied quality. However, the quality of information disseminated in the field of medicine has been questioned as the negative health consequences of health misinformation can be life threatening. There is currently no generic automated tool for evaluating the quality of online health information spanned over broad range. To address this gap, in this paper, we applied data mining approach to automatically assess the quality of online health articles based on $10$ quality criteria. We have prepared a labelled dataset with $53012$ features and applied different feature selection methods to identify the best feature subset with which our trained classifier achieved an accuracy of $84\%-90\%$ varied over $10$ criteria. Our semantic analysis of features shows the underpinning associations between the selected features \& assessment criteria and further rationalize our assessment approach.
Our findings will help in identifying high quality health articles and thus aiding users in shaping their opinion to make right choice while picking health related help from online. 
\end{abstract}


%
\IEEEpeerreviewmaketitle

\IEEEdisplaynontitleabstractindextext
\ifCLASSOPTIONcompsoc
\IEEEraisesectionheading{\section{Introduction}\label{sec:introduction}}
\else
\section{Introduction}
\label{sec:introduction}
\fi
\IEEEPARstart{T}{he} tremendous advancement of digital technology and widespread usage of Internet have made information accessible worldwide. Consequently, majority of people are turning to the Internet for searching a diverse range of health related information. 
According to a study by Australian Institute of Health and Welfare, 78\% of Australian adults were found to search health-related information in 2015 \cite{AustralianInstituteofHealthandWelfare}. However, the reliability of information from web sources are questionable due to the unregulated nature of Internet. 


 In this era of Internet, misinformation (dubious, low quality fabricated information) disseminates much faster like wildfire than the truth. A plethora of information from online health articles (OHA) and other sources (Blogs, Facebook, Twitter, YouTube, etc.) are available for health information quester. But all the information are not reliable as these stem from various individuals and organization~\cite{Sbaffi2017, Kitchens2014, Eysenbach2002}. Hence, the task of distinguishing unreliable health information from reliable one poses substantial challenges on individuals \cite{Bhatt2017}, \cite{in1}, \cite{in2}. 

The extensive spread of unreliable information can negatively affect public health. Misinformation based wrong decision forces people to uphold erroneous belief and opinions instead of irrefutable evidence \cite{Shu2017}. Sometimes, these types of articles fail to render  envisioned  information to pursuer. This may result misinterpretation of concept and eventually trigger fear and incite one to change regular habit overnight. However, the online network isn't going anywhere and seeking and sharing health information online will not be stopped. Misinformation will prevail as well \cite{l15}. For this reason, assessing and assuring the quality of health information on World Wide Web becomes a fundamental issue for users \cite{Eysenbach}. The better the quality of health information, the more reliable and accessible it is and the more effective it will be in moulding user’s behaviour towards health-care.

 In order to curb this situation, several approaches have been proposed to assess the quality of health related information. Among these, some of the approaches conducted assessment manually and demanded user’s perception to qualify a health news. A number of studies estimated the quality of the overall web sources rather evaluating each of the article published in it~\cite{l18,new1}. A few others tried to evaluate the quality of articles published in specific disease domain which narrowed down the scope of their work \cite{l11}, \cite{l41}, \cite{l42}. Some studies proposed evaluation criteria framework and some tried to assess quality based on that proposed framework \cite{l10,l15}. But in case of criteria selection, a question is always there about its specific application on medical domain as criteria selection for health specific articles necessitate the involvement of health professionals.  However, given the ever changing landscape of Internet, no universal framework for automatically assessing the quality of OHA has been proposed to date. With this context in mind, this study attempts to automate the quality assessment process of OHA based on the ideas and effort from HealthNewsReview.org\footnote{\url{https://www.healthnewsreview.org}}. This organization manually evaluates health-related articles by a team comprised 50 experts from various disciplines including journalism, medicine, health services research, public health and patient perspectives. Performance of this organization is excellent but not scalable in comparison to the speed of information explosion worldwide. In this paper, we applied a data mining based approach to assess the quality of online health articles automatically. Our main contributions can be summarized as follows:
 
\begin{itemize}
    \item We have developed a labelled dataset of health related news articles which were finely annotated by health experts from HealthNewsReview.org. So far, no generic health related dataset is available that is suitable for assessing the quality of OHA. Our dataset, once released, will be a valuable resource for health and research communities for conducting future studies on the topic of misinformation in the field of medicine.
    \item We have explored multifaceted feature spaces through systematic content analysis to identify appropriate features to automate quality assessment process. We have also keyed out criteria-wise discriminating features by analyzing feature importance.
    \item We have examined the applicability of various data mining techniques in assessing the quality of OHA automatically and  achieved state-of-the-art performance on it.
    \item We also have provided explanation of feature subset corresponding to each criterion to justify the value of the assessment.
\end{itemize}

\section{Related Work}
\label{sec-rw}

Quality of the online health related information has been a major concern from the dawn of the World Wide Web (WWW) era \cite{l27}, \cite{l28}. 
Numerous tools have been developed to alleviate the quality measurement of health related information most of which are based on a particular disease (e.g., cancer, diabetes, etc.) and lack in robust validity and reliability testing. In \cite{l5}, Keselman et. al. conducted an exploratory study with a view to developing a methodological approach to analyze health related web pages and apply it to a set of relevant web pages. This qualitative study analysed webpages about natural treatment of diabetes to accentuate the challenges faced by consumers in seeking health information. It has also underscored the importance of developing support tools so that this formative study could help users to seek, evaluate, and analyze information in the changing digital ecosystem.

We have summarized the relevant research along three categories. The first is characterizing quality assessment tools built on inter-rater agreements by expert panels. These studies aimed at judging the quality of written consumer health information \cite{le}, \cite{l2}, \cite{l3}, \cite{l4}, \cite{l34} and health reports in lay media \cite{l36}, \cite{l37}, \cite{l29}, \cite{l38}, \cite{l1}, \cite{l40}. The second is characterizing quality assessment approaches built on a checklist of factors. These studies focused on identifying appropriate criteria list for evaluating online health information \cite{l30}, \cite{l25},\cite{l31}, \cite{l15}, \cite{Eysenbach}, \cite{l6}, \cite{l32}, \cite{l7}, \cite{l11}, \cite{l12},\cite{l10}, \cite{l33}. And the third one is characterizing the approaches built on machine learning techniques to automate health information related analysis including health dataset preparation \cite{l19}, improving veracity of medical information \cite{l16}, tracking misinformation for specific disease domain~\cite{l17}, \cite{l35} and reliable health media structure analysis \cite{l18}.

\subsection{Statistical Analysis Based Quality Assessment Approach}
\textit{DISCERN}~\cite{l2}, a short instrument, was developed for judging the quality of written consumer health information about treatment choices by producers, health professionals and patients, and for facilitating the production of high quality evidence-based patient information. The DISCERN approach was a combination of qualitative methods and a statistical measure of inter-rater agreements among expert panel representing a range of expertise  including production and use of consumer health information~\cite{le}. For establishing the face and content validity, and inter-rater reliability, this approach administered questionnaire to information providers and self-help organizations. Later, authors of \cite{l2} developed an explicit scheme for calculating a 5-star quality rating system for consumer health information based on DISCERN \cite{l3}. 

The \textit{Ensuring Quality Information for Patients (EQIP)}~\cite{l4} is another tool to assess the presentation quality of all types of written health care information in a more rigorous way, and to prescribe the action that is required following the evaluation. EQIP tool was demonstrated through several processes of item generation, testing for concurrent validity, inter‐rater reliability and utility using large diverse samples of written health care information.

The \textit{Quality Index for health related Media Reports (QIMR)} was developed as an evaluation tool to monitor the quality of health research reporting in the lay media, more specifically for Canadian media. Themes from interviews with health journalists and researchers were undertaken to develop QIMR \cite{l1}. However, QIMR approach is limited in sample size and scope, and failed to evaluate quality of news sources having content of varying quality. 

However, specific focus on treatment information or particular media has narrowed down the scope of these approaches on different applications and questions their applicability to online content about other aspects of health and illness. On the contrary, our approach is applicable to all health related information domains. Moreover, the existing approaches were conducted through manual labour, whereas ours is fully automated system to assess quality of health articles in a shorter possible time.

\subsection{Criteria Based Quality Assessment Approach}

To date, there is no clear universal standard to assess the quality of web based health information \cite{l25}. Kim et. al. conducted extensive review to identify criteria that were already proposed or employed specifically for evaluating health related information world wide \cite{l26}. Eysenbach et. al. conducted a systematic review to compile criteria actually used to measure the quality of health information on the Web and  synthesized evaluation results from studies containing quantitative data on structure and process \cite{Eysenbach}.  Comparing the methodological frameworks of existing approaches authors concluded with the need for defining operational criteria for quality assessment. \cite {Sbaffi2017} is another systematic review where authors reviewed empirical studies on trust and credibility in the use of web-based health information (WHI) with an aim to identify factors that impact judgments of trustworthiness and credibility, and to explore the role of demographic factors affecting trust formation.

The Code of Conduct for medical websites (HONcode), initiated by the Health On the Net Foundation, was the first attempt to propose guidelines to information providers for raising the quality of medical and health information available on the World Wide Web \cite{l6}. Adopting a set of eight criteria to certify websites containing health information, its creators also developed a Health Website Evaluation Tool, which offered users with an indication of commitment to quality from the providers. 

There are several criteria-based assessment tools and few of them have proper validation \cite{l7}. \textit{Quality Evaluation Scoring Tool (QUEST)} is the first quantitative tool that supports a broad range of health information and had undergone a validation process~\cite{l10}. Based on a review of existing tools \cite{l11,l12}, QUEST quantitatively measures six criteria: authorship, attribution, conflicts of interest, currency, complementarity and tone which can be used by health care professionals and researchers alike. QUEST’s reliability and validity were demonstrated by evaluating online articles on Alzheimer’s disease . In an Fuzzy VIKOR based approach, Afful-Dadzie et. al.~\cite{l15} proposed a new criteria framework for measuring the quality of information provided by each site. Authors demonstrated a decision making model to find out how online health information providers could be assessed and ranked based on their quality.

However, some proposed criteria across existing literature are specific for particular domain while some are common which can be considered for standardizing universal set of criteria. In \textit{HealthNewsReview.org}, a group of expert reviewed ten criteria based on analysis from previous studies combined with viewpoint from health care journalism\footnote{\url{https://healthjournalism.org/secondarypage-details.php?id=56}}.Basic issues that a consumer should know for developing their opinions on health related interventions were addressed by these ten criteria. Characteristics of ten defined criteria from standards of health reporting perspectives and all possible basic points with a view to serve the interests of the public have convinced us to adopt these set as standard for evaluating health related articles.
 
\subsection{Machine Learning Based Analysis and Miscellaneous}

Apart from aforementioned approaches, there are few more studies which are not directly aligned with our research but provide us with valuable insights.

In \cite{l19}, authors developed a new labelled dataset of misinformative and non-misinformative comments from a medical health forum, MedHelp, with a view to making a resource for medical research communities to study the spread of medical misinformation. Preliminary feature analysis of the dataset was also presented to develop a real-time automated system for identifying and classifying medical misinformation in online forums.

An applied machine learning based approach is proposed in \cite{l16}, where authors addressed the veracity of online health information by automating systemic approaches in conjunction with Evidence-Based Medicine (EBM). Based on EBM and trusted medical information sources, authors proposed an algorithm, MedFact, which would recommend trusted medical information within health related social media and empower online users to determine the veracity of health information using machine learning techniques. Their aim was to address the factual accuracy of online health information from social media discourse based on keyword extraction. Whereas, our objective is to evaluate the quality of online health realted articles from datamining perspective. we have focused on identifying the discriminating features of health related articles for assessing the quality in a automatic manner.

Ghenai et. al. \cite{l17} proposed a tool for tracking misinformation around health concerns on Twitter based on a case study about Zika. The tool discovered health related rumours in social media by incorporating professional health experts through crowdsourcing for annotating dataset and machine learning for rumour classification. Our aim is different from this study. Rather than focusing on health related rumour, we focused on all types of health related articles available online to evaluate their quality so that people could be able to identify which articles to read or which to avoid for decision making. 

A recent study by Dhoju et. al.~\cite{l18} has identified structural, topical and semantic differences between health related news articles from reliable and unreliable media by conducting a systematic content analysis. By leveraging a large-scale dataset, authors successfully identified some discriminating features which separate reliable health news from the unreliable one. 


However, our study is quite different from these already existing methodologies. Our work is based on the initiatives of HealthNewsReview.org, which aimed to evaluate the quality of health related articles as soon as they appeared in the news media in a manual process. Keeping new information arrival rate in mind, our aim is to automate this quality assessment process from data mining perspective, which has not been examined to date according to our knowledge. Our goal is to use their finely tuned manually annotated health information in our study to examine the performance of automated quality assessment approach in health domain. This organization proposed ten criteria as a standard of judging the quality of health articles. Though various criteria framework have been proposed in literature for assessing health information quality, criteria list proposed by HealthNewsReview.org was more standardized. Considering existing frameworks and experts opinion regarding quality of health information, this organization aimed to address the basic issues of health interventions through ten criteria so that consumers could develop informed opinions about these interventions and how/whether they matter in their lives. Our objective is to address each individual criteria from data mining perspective and discover to what degree each criterion can be automated.

\section{Dataset Description}
\label{data-prep}
There is currently no single dataset for assessing the quality of online health articles (OHA). For this study, we have prepared a dataset based on 1720 health-related articles from HealthNewsReview.org.
The mission of this website is to introduce a significant step towards meaningful health care reform by evaluating the accuracy of medical news and examining the quality of evidence they provide. Since its foundation in 2006, HealthNewsReview.org provides reviews of health news reporting from major U.S. news organizations conducted by a multi-disciplinary team of reviewers from journalism, medicine, health services research and public health domain.

According to the editorial team of HealthNewsReview.org, all stories and press news releases about public health interventions should be evaluated by \textit{ten} different criteria to ensure the quality of information in terms of accuracy, balance and completeness. Reviewers justified each of the criterion with `satisfactory' or `Not Satisfactory' scores based on their quality. In some cases, some criteria were rated as `Not Applicable' where it was impossible or unreasonable for an article to address those. Below we provide a list of those criteria\footnote{\url{https://www.healthnewsreview.org/about-us/review-criteria/}}.\newline
\begin{description}[style=multiline,leftmargin=2.2cm,font=\normalfont]
  \item[$\bullet$ Criterion 1] Does the story adequately discuss the costs of the intervention?
  \item[$\bullet$ Criterion 2] Does the story adequately quantify the benefits of the intervention?
  \item[$\bullet$ Criterion 3] Does the story adequately explain/quantify the harms of the intervention?
  \item[$\bullet$ Criterion 4] Does the story seem to grasp the quality of the evidence?
  \item[$\bullet$ Criterion 5] Does the story commit disease-mongering?
  \item[$\bullet$ Criterion 6] Does the story use independent sources and identify conflicts of interest?
  \item[$\bullet$ Criterion 7] Does the story compare the new approach with existing alternatives?
  \item[$\bullet$ Criterion 8] Does the story establish the availability of the treatment/test/product/procedure?
  \item[$\bullet$ Criterion 9] Does the story establish the true novelty of the approach?
  \item[$\bullet$ Criterion 10] Does the story appear to rely solely or largely on a news release?
\end{description}

Based on these ten specific criteria the reviewers have graded stories which are about surgical procedures, drugs or devices, dietary recommendations, vitamins or nutritional supplements, diagnostic and screening tests and psychotherapy/ mental health interventions. We have considered these reviews as gold standard records for our approach.

\subsection{Data Collection}
\label{subsec-data-collection}
To collect data from HelathNewsReview.org we have created GUI app using C\#.Net framework. Since the website has no API, we have created our scraper using HTML Agility Pack\footnote{\url{https://html-agility-pack.net}}, a free and open source tool to extract data from website, and stored data in MS SQL database. For each review, we gathered title of the original news, corresponding link of original news, category and criteria wise score as `Satisfactory', `Unsatisfactory' or `Not Applicable'. We have collected all reviewed stories from 2006 to 2018 and reviewed press news releases from 2015 to 2018 from the website and removed duplicity as same story may coexist under different categories. The source URL is then accessed using Newspaper3k\footnote{\url{https://newspaper.readthedocs.io/en/latest/}}, a python3 library for extracting and curating articles.

\begin{figure}[!th]    
    \centering
    \includegraphics[width=\linewidth]{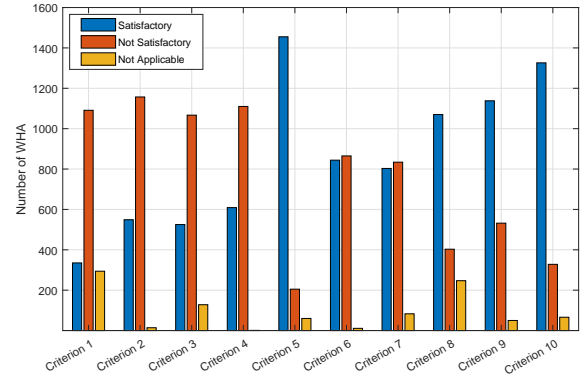}
    \caption{Criteria wise class distribution over entire corpus.}
    \label{criteria_distribution}
\end{figure}

Overall, our dataset consists of three class labels: Satisfactory, Not Satisfactory and Not Applicable, for each of the ten criteria. Figure \ref{criteria_distribution} shows the criteria wise distribution of class labels over 1720 data corpus. 

As we can see that the number of observations belonging to ‘Not Applicable’ class is significantly lower than that of other two classes in every criteria, we have omitted this class value for our initial study.

\section{Feature Engineering}
\label{sec-QA-approch}
In this section, we will explain the data pre-processing, feature extraction and feature selection process to establish baseline performance for our approach. All our data pre-processing and feature extraction have been conducted using python, and some other useful library, e.g., scikit-learn\footnote{\url{https://scikit-learn.org/stable}} and NLTK\footnote{\url{https://www.nltk.org}}.

\subsection{Data Pre-Processing}
\label{subsec-data process}
Certain refinement of raw data is essential for removing irrelevant information and reducing the size of actual data \cite{l46}. To enhance the accuracy and performance of our classification model, we have run step by step data pre-processing tasks on each article. 

\begin{figure}[th!]
 \centering
     \includegraphics[scale=.5]{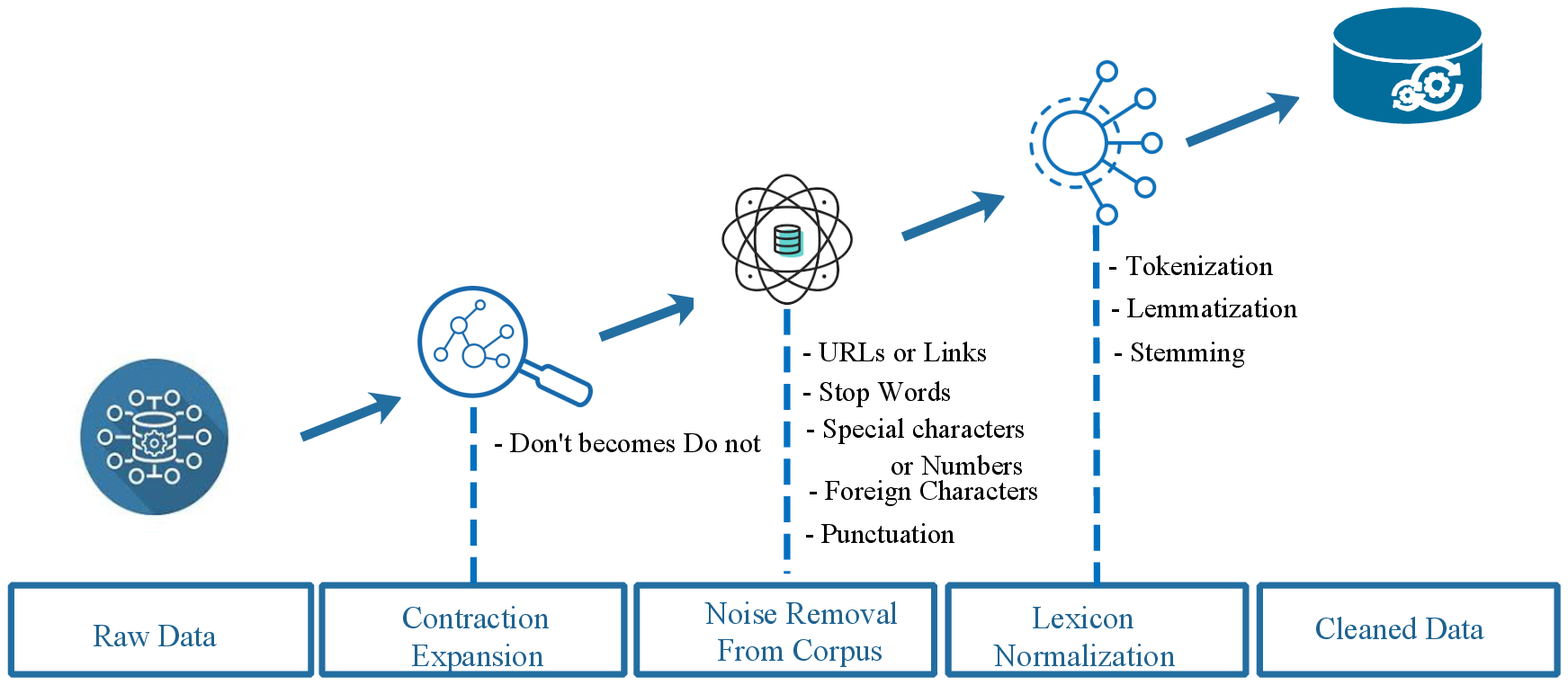}
    \caption{Data pre-processing pipeline.}
    \label{pipeline}
\end{figure}

Figure \ref{pipeline} shows the text pre-processing pipeline for cleaning our data. Following three refinement steps are adopted:


\subsubsection{Contraction Expansion}
Contractions are shortened version of words or syllables which pose problems in text analytics. To help text standardization with original form of words, each contraction has been expanded to its main form. For instance, expanded form of `i'd' and `you've' became `I would' and `you have' respectively.
\subsubsection{Noise Removal}
 Noise removal is one of the most important text pre-processing steps. Usually, URLs, special characters and symbols add extra noise in unstructured text. We applied punctuation removal, special character removal, html formatting removal and numbers removal to get rid of these noise. Because of having little significance in corpus, we removed stop words (words like: a, the, is, me, etc.) as well.
\subsubsection{Word Normalization}
In text analytics, tokenization of document is required for identifying meaningful keywords. Apart from tokenizing documents, stemming and lemmatization have also been used for reducing inflectional forms of word (connet, connected, connection, etc.) and derivationaly related forms of word to a common base form. 

Remaining chunks of cleaned text data are then fed for feature extraction.

\subsection{Feature Extraction}
\label{sec-feature-development}
Multiple categories of features have been extracted for classifying criteria. For model construction, we have keyed out some features which might help in prediction of classes. The complete set of extracted features with their corresponding description is depicted in Table \ref{tab:ft}.

{\renewcommand{\arraystretch}{1}{\begin{table*}
\caption{List of extracted features with brief description}
\label{tab:ft}
\begin{tabular}{|p{3.7cm}|p{3cm}|p{7.7cm}|p{.7cm}|p{0.9cm}|}\hline
Scope & Feature Name& Description  & Feature Number & Output Type \\ \hline\hline
Linguistic measure                            & LIWC                           & Measures textual features                                                                          & 93             & Real        \\ \hline
Word frequency                                & TF-IDF                         & Measures the importance of a word in a document                                                    & 4000           & Real        \\ \hline
\multirow{2}{*}{Word-category disambiguation} & POS Tag                        & Counts the number of part of speech in a document                                                  & 35             & Integer     \\ \cline{2-5} 
                                              & POS Word                       & Defines the parts of speech of each word separately and then counts its number within the document & 47450         & Integer     \\ \hline
\multirow{3}{*}{Citation and Ranking}         & Internal Links                 & Defines the number of self-citations                                                               & 1              & Integer     \\ \cline{2-5} 
                                              & External Links                 & Defines the number of external citations                                                           & 1              & Integer     \\ \cline{2-5} 
                                              & Alexa Rating                   & Ranks every document having link according to alexa rating                                         & 1428           & Real        \\ \hline
Similarity measure                            & Cosine similarity              & Measures the relation between headline and body                                                    & 1              & Real        \\ \hline
\multirow{3}{*}{Miscellaneous}                & Normalized distinct word count & Measures how many distinct words were used in the text                                             & 1              & Real        \\ \cline{2-5} 
                                              & Per\_num\_count                & Counts the number of person mentioned in the text                                                  & 1              & Integer     \\ \cline{2-5} 
                                              & Org\_num\_count                & Counts the number of organization mentioned in the text                                            & 1              & Integer     \\ \hline
\end{tabular}
\end{table*}
    }

We have considered the following features:
\subsubsection{Linguistic Inquiry and Word Count (LIWC)}
\label{subse-liwc}
To obtain a wide variety of psychological and linguistic features, we apply LIWC2015 \cite{liwc}, a transparent text analysis program to score words in psychologically meaningful categories, on original news texts in our dataset. LIWC calculates the following dimensions:
 \vspace*{-2mm}
\begin{itemize}
    \item Summary Dimension (Consists of 8 features; e.g., word count, word per sentence)
\item Punctuation mark (Consists of 12 features; e.g., comma, colon, quote)
\item Function words (Consists of 15 features; e.g., Pronoun, article, conjunction)
\item Perceptual process (Consists of 4 features; e.g., see, hear)
\item Biological process (Consists of 5 features; e.g., Body, health)
\item Drives (Consists of 6 features; e.g., reward, risk, power)
\item Other grammar (Consists of 6 features; e.g., interrogatives, numbers)
\item Time orientation (Consists of 3 features; e.g., past, present, future)
\item Relativity (Consists of 4 features; e.g., motion, time)
\item Affect (Consists of 6 features; positive emotion, negative emotion (e.g. anger))
\item Personal concerns (Consists of 6 features; e.g., word, leisure, money)
\item Social (Consists of 5 features; e.g., Family, friend)
\item Informal language (Consists of 6 features; e.g., filler, swear)
\item Cognitive process (Consists of 7 features; e.g., Differ, Insight)
\end{itemize}

\subsubsection{Term Frequency and Inverse Document Frequency}
\label{subsec-tf-idf}
We have used this weighting metric to measure the importance of a term in a document within entire dataset \cite{l46}. Term Frequency (TF) is used to quantify the frequency of a word in a particular document. On the contrary, Inverse Document Frequency (IDF) measures the importance of a term within the corpus. 
Let, $D$ symbolizes the whole corpus with $N$ documents. If $n(t)_d$ denotes the number of times term $t$ appears in a document $d$, then TF, denoted by $TF(t)_d$, can be calculated by equation \eqref{eq:1}:
 \vspace*{-3mm}
\begin{align} \label{eq:1}
TF(t)_d  = \frac{n(t)_d}{\sum_{\acute{t} \in d} n(\acute{t})_d},
\end{align}

And IDF, denoted by $IDF(t)_D$, can be calculated by equation \eqref{eq:2}:
 \vspace*{-2mm}
\begin{align} \label{eq:2}
IDF(t)_D = 1 + \log[N \times |{\{d \in D \colon t \in d}\}|^{-1}],
\end{align}

For a particular term, the product of TF and IDF represents the TF-IDF weight of that term. The higher the TF-IDF score, the rarer the term is. We applied unigram tokenizer on our text and eliminated features with extremely low as well as extremely high frequency for achieving a better accuracy~\cite{tf1}~\cite{tf2}. Since too frequent or too rare words are not influential in characterizing an article, we ignored all words that have appeared in more than 90\% of the documents and less than 3 documents. Again, to keep the dimensionality of our feature set to a manageable size, we set maximum feature count to top 4000 terms based on frequency.

\subsubsection{Part Of Speech Tagging}
\label{subsec-link}
Part Of Speech Tagging (POST), also known as word-category disambiguation, is used to annotate word with appropriate part-of-speech based on both its definition and context to resolve lexical ambiguity \cite{post}.  To recognize POST, we have applied Stanford postagger\footnote{\url{https://nlp.stanford.edu/software/tagger.html}}. We found 35 tagset (list of part-of-speech tags, e.g., CC, CD, NP, RBR, etc. ) in the corpus with which we derived two sets of features: POS tag count and POSWord count. For POS tag count, we measured the document wise count of words belonging to a particular POS tag and thus found 35 individual features. On the other hand, for POSWord count, we measured the count of tag associated with each individual word within a document and found 47,451 non-overlapping features. POSWord features are capable of performing rudimentary word sense disambiguation in situations where a word can represent several meanings. 

\subsubsection{Citation and Ranking}
\label{subsec-link}
We analysed the presence of hyperlinks to determine the credibility of an article. We extracted three features -- \textit{internal link}, \textit{external link} and \textit{Rank} from link attribute. We counted the number of internal links to inspect the amount of self-citation occurred in a document so that we can predict some biasness in it. Conversely, number of external links were counted to predict the citation network of an article. We derived the \textit{rank} attribute to envisage the quality of the article by measuring the superiority of the webpages that particular article cited to. We considered Alexa Global Ranking\footnote{\url{https://www.alexa.com/siteinfo}} as an indicator of superiority measurement of a webpage as it gives an estimation of a website’s popularity. We counted outgoing links from all the documents within the corpus and found 1428 distinct domains. We replaced each domain with its associated alexa rank value and thus, we found 1428 distinct rank features for overall corpus.
\subsubsection{Similarity Measure}
\label{subsec-Similarity measure}
Ambiguous and misleading headline can degrade the quality of an article. So, to measure the relevance between headline-body pair of each article, we used TF-IDF Cosine similarity metric to extract similarity feature \cite{coss}. It quantifies the similarity between headline and body of the document irrespective of their size by measuring the cosine of the angle between two vectors projected in a multi-dimensional space. 
\subsubsection{Miscellaneous}
\label{subsec-other}
We quantified \textit{normalized distinct word count} as a feature to determine how rare a word contributes in the classification problem as health related articles comprise different medical terms. We have also counted the number of organizations and person mentioned in articles to predict biasness. We used Stanford Named Entity Recognizer (NER)\footnote{\url{https://nlp.stanford.edu/software/CRF-NER.html}} to extract these features.

\subsection{Feature Selection}
\label{sec-feature-selection}
We are aiming to predict ten different criteria using numerous features (total of $53012$) some of which might redundant or irrelevant to make predictions. Dataset containing irrelevant features can result in over-fitting. It also can mislead the modelling power of a method. Thus, it is critically important to select most relevant features from the feature set. In order to select the features that contribute most in our classification task, we have employed three different automatic feature selection techniques. First, correlation-based attribute evaluation ($C_oAE-PC$), which evaluates worth of a feature by measuring Pearson’s correlation between it and the class. Second, Classifier-based attribute evaluation ($C_lAE-LR$), which evaluates the worth of a feature using Logistic Regression classifier. Third, Classifier-based attribute evaluation ($C_lAE-RF$), which evaluates the worth of a feature using Random Forest classifier. For each of the above three attribute evaluator, rank search method was performed which ranks features by their individual evaluations to find out the most correlated feature set.
\section{Experimental Evaluation}
\label{sec-evaluation}

The core contribution of our work is to assess the quality of online health articles automatically 
applying various data mining techniques. In this section, we have quantified and evaluated the performance of a number of classification techniques, for different feature selection methods and variable feature sizes, to achieve the best result.

\subsection{Evaluate Classification Techniques}
We have experimented four prominent classification techniques on our dataset and reported their results. We have performed a binary class (Satisfactory and Not Satisfactory) classification using three supervised learning methods and one ensemble method for obtaining better accuracy in assessing quality of OHA. First, Support Vector Machine (SVM) algorithm, which uses kernel trick to implicitly mapping inputs into a high-dimensional feature space \cite{10}. We have used PolyKernel as kernel to control the projection and the amount of flexibility in separating classes in our dataset. Second, Naive Bayes classification algorithm which calculates the posterior probability for each class using a simple implementation of the Bayes theorem and makes the prediction for the class with the highest probability. For each numerical attribute, a Gaussian distribution is assumed by default \cite{12}. Third, Random Forest classifier which constructs a multitude of decision trees at training time and merges them together to get a more accurate and stable prediction \cite{11}. We considered $100$ trees for Random Forest implementation. Forth, EnsembleVoteClassifier, a meta-classifier combining similar or conceptually different machine learning classifiers for classification via majority voting. We have combined three aforementioned classifiers to build our ensemble estimator and examined its performance on our dataset.
All methods were evaluated by 10-fold cross-validation, where in each validation $90\%$ of dataset was used for training purpose and $10\%$ for testing. Various combinations of the extracted features have been experimented to evaluate how accurately our approach can automatically classify each criterion.
\subsection{Identify Feature Selection Method and Feature Size}
To identify the feature selection method and the feature size that result best classification accuracy for our dataset, We have experimented the impact of different feature selection methods and varied feature sizes on classification accuracy.

\subsubsection{Identify Feature Selection Method}
We ran three feature selection methods on our feature space, with a goal of determining which feature selection method performs best by selecting a best feature subset that results best classification performance.
Table \ref{tab:fs} presents the outcomes of the comparative study of three different feature selection methods over four different classifiers (SVM, Naive Bayes, Random Forest and EnsembleVote) carried out against a feature subset with feature size $4000$. Here, we have presented weighted Precision ($W_P$), weighted Recall ($W_R$) and weighted F-Measure ($W_F$) from the Weka~\cite{Zeng2014} output for presenting a better estimate of overall classification performance. Weka calculates weighted average by taking average of each class, weighted by the proportion of how many elements are in each class. So, for our binary class problem, $W_P$, $W_R$ and $W_F$ are calculated from equation \eqref{eq:wp}, \eqref{eq:wr} and \eqref{eq:wf} respectively.
\setlength{\belowdisplayskip}{0pt} \setlength{\belowdisplayshortskip}{0pt}
\setlength{\abovedisplayskip}{0pt} \setlength{\abovedisplayshortskip}{0pt}
\begin{align} 
\label{eq:wp}
W_P  = \frac{(P_{CS}\times \left|CS\right|)+(P_{CNS}\times \left|CNS\right|)}{\left|CS\right|+\left|CNS\right|},
\end{align}
\begin{align} 
\label{eq:wr}
W_R  = \frac{(R_{CS}\times \left|CS\right|)+(R_{CNS}\times \left|CNS\right|)}{\left|CS\right|+\left|CNS\right|},
\end{align}
\begin{align} 
\label{eq:wf}
W_F  = \frac{(F_{CS}\times \left|CS\right|)+(F_{CNS}\times \left|CNS\right|)}{\left|CS\right|+\left|CNS\right|},
\end{align}
Where, $P_{CS}$ and $P_{CNS}$ are the Precisions for class `Satisfactory' and `Not Satisfactory'; 
$R_{CS}$ and $R_{CNS}$ are the Recalls for class `Satisfactory' and `Not Satisfactory'; $F_{CS}$ and $F_{CNS}$ are the F-Measures for class `Satisfactory' and `Not Satisfactory'; 
$\left|CS\right|$ and $\left|CNS\right|$ are the number of instances in class `Satisfactory' and `Not Satisfactory' respectively.

For all criteria, clearly SVM performed well among all four classifiers. We also observe from the table \ref{tab:fs} that, for all criteria (except criterion 3 and criterion 5), the Pearson's correlation feature selection method ($C_oAE-PC$) performs best for the SVM classifer. For criteria 3 and 5, the Logistic Regression feature selection performs slightly better than the Pearson's correlation method.

{\renewcommand{\arraystretch}{0.8}
\begin{table*}[]
\caption{Comparison study of three feature selection methods over four classifiers (Feature Size: $4000$)}
\begin{tabular}{|p{1cm}|p{1.55cm}|p{.8cm}|p{.8cm}|p{.8cm}|p{.8cm}|p{.8cm}|p{.8cm}|p{.8cm}|p{.8cm}|p{.8cm}|p{.8cm}|p{.8cm}|p{.8cm}|} \hline
\multirow{2}{*}{Criterion} & \multirow{2}{*}{FS Methods}    & \multicolumn{3}{c|}{SVM}   & \multicolumn{3}{c|}{Random Forest} & \multicolumn{3}{c|}{Naive Bayes}                     & \multicolumn{3}{c|}{Ensemble}\\ \cline{3-14} 
          &               &\multicolumn{1}{c|} {$W_P$}     & \multicolumn{1}{c|}{$W_R$}             & \multicolumn{1}{c|}{$W_F$}                               &  \multicolumn{1}{c|}{$W_P$}        &  \multicolumn{1}{c|}{$W_R$}     &  \multicolumn{1}{c|}{$W_F$}     &  \multicolumn{1}{c|}{$W_P$}     & \multicolumn{1}{c|}{$W_R$}     & \multicolumn{1}{c|}{$W_F$}     & \multicolumn{1}{c|}{$W_P$}     & \multicolumn{1}{c|}{$W_R$}     &\multicolumn{1}{c|}{$W_F$}      \\ \hline\hline
1         & $C_{o}AE-PC$ & 0.901 & 0.903         & \textbf{0.899} & 0.794    & 0.786 & 0.718 & 0.861 & 0.756 & 0.774 & 0.868 & 0.803 & 0.816  \\
          & $C_{l}AE-LR$ & 0.887 & 0.888         & 0.881                           & 0.826    & 0.786 & 0.711 & 0.813 & 0.684 & 0.708 & 0.859 & 0.827 & 0.836  \\
          & $C_{l}AE-RF$ & 0.877 & 0.881         & 0.875                           & 0.831    & 0.792 & 0.724 & 0.792 & 0.637 & 0.666 & 0.848 & 0.813 & 0.823  \\ \hline\hline
2         & $C_{o}AE-PC$ & 0.855 & 0.857         & \textbf{0.854} & 0.739    & 0.707 & 0.621 & 0.793 & 0.721 & 0.730 & 0.799 & 0.738 & 0.747  \\
          & $C_{l}AE-LR$ & 0.826 & 0.828         & 0.821                           & 0.765    & 0.707 & 0.615 & 0.766 & 0.686 & 0.696 & 0.789 & 0.739 & 0.747  \\
          & $C_{l}AE-RF$ & 0.795 & 0.801         & 0.794                           & 0.751    & 0.703 & 0.610 & 0.691 & 0.567 & 0.587 & 0.729 & 0.674 & 0.685  \\ \hline\hline
3         & $C_{o}AE-PC$ & 0.835 & 0.837         & 0.832                           & 0.752    & 0.720 & 0.655 & 0.790 & 0.712 & 0.720 & 0.793 & 0.727 & 0.735  \\
          & $C_{l}AE-LR$ & 0.851 & 0.849         & \textbf{0.841} & 0.779    & 0.722 & 0.651 & 0.764 & 0.698 & 0.707 & 0.773 & 0.727 & 0.735  \\
          & $C_{l}AE-RF$ & 0.804 & 0.808         & 0.803                           & 0.780    & 0.721 & 0.648 & 0.704 & 0.612 & 0.624 & 0.729 & 0.674 & 0.685  \\ \hline\hline
4         & $C_{o}AE-PC$ & 0.847 & 0.848         & \textbf{0.846} & 0.707    & 0.695 & 0.635 & 0.783 & 0.713 & 0.718 & 0.790 & 0.728 & 0.733  \\
          & $C_{l}AE-LR$ & 0.824 & 0.824         & 0.818                           & 0.746    & 0.693 & 0.615 & 0.758 & 0.689 & 0.694 & 0.769 & 0.720 & 0.726  \\
          & $C_{l}AE-RF$ & 0.779 & 0.783         & 0.778                           & 0.741    & 0.692 & 0.615 & 0.680 & 0.583 & 0.592 & 0.704 & 0.643 & 0.650  \\\hline\hline
5         & $C_{o}AE-PC$ & 0.894 & 0.887         & 0.843                           & 0.769    & 0.877 & 0.819 & 0.864 & 0.722 & 0.767 & 0.873 & 0.890 & 0.863. \\
          & $C_{l}AE-LR$ & 0.888 & 0.901         & \textbf{0.888} & 0.769    & 0.877 & 0.819 & 0.829 & 0.757 & 0.786 & 0.869 & 0.887 & 0.873  \\
          & $C_{l}AE-RF$ & 0.856 & 0.880         & 0.862                           & 0.892    & 0.877 & 0.820 & 0.805 & 0.707 & 0.748 & 0.845 & 0.873 & 0.852  \\ \hline\hline
6         & $C_{o}AE-PC$ & 0.835 & 0.835         & \textbf{0.835} & 0.688    & 0.688 & 0.688 & 0.754 & 0.744 & 0.742 & 0.747 & 0.737 & 0.735  \\
          & $C_{l}AE-LR$ & 0.733 & 0.733         & 0.732                           & 0.678    & 0.676 & 0.674 & 0.689 & 0.644 & 0.623 & 0.693 & 0.648 & 0.628  \\
          & $C_{l}AE-RF$ & 0.719 & 0.719         & 0.719                           & 0.687    & 0.686 & 0.685 & 0.665 & 0.636 & 0.621 & 0.667 & 0.638 & 0.624  \\\hline\hline
7         & $C_{o}AE-PC$ & 0.855 & 0.854         & \textbf{0.854} & 0.669    & 0.666 & 0.663 & 0.737 & 0.733 & 0.732 & 0.747 & 0.737 & 0.735  \\
          & $C_{l}AE-LR$ & 0.716 & 0.715         & 0.715                           & 0.678    & 0.666 & 0.659 & 0.669 & 0.630 & 0.611 & 0.669 & 0.630 & 0.611  \\
          & $C_{l}AE-RF$ & 0.690 & 0.689         & 0.688                           & 0.644    & 0.638 & 0.632 & 0.627 & 0.607 & 0.595 & 0.627 & 0.607 & 0.595  \\\hline\hline
8         & $C_{o}AE-PC$ & 0.875 & 0.876         & \textbf{0.870} & 0.768    & 0.737 & 0.638 & 0.813 & 0.777 & 0.786 & 0.813 & 0.781 & 0.789  \\
          & $C_{l}AE-LR$ & 0.814 & 0.821         & 0.807                           & 0.780    & 0.737 & 0.638 & 0.710 & 0.642 & 0.661 & 0.737 & 0.712 & 0.721  \\
          & $C_{l}AE-RF$ & 0.762 & 0.775         & 0.765                           & 0.750    & 0.736 & 0.639 & 0.657 & 0.523 & 0.563 & 0.718 & 0.699 & 0.706  \\\hline\hline
9         & $C_{o}AE-PC$ & 0.867 & 0.869         & \textbf{0.867} & 0.695    & 0.698 & 0.607 & 0.782 & 0.765 & 0.770 & 0.782 & 0.765 & 0.770  \\
          & $C_{l}AE-LR$ & 0.827 & 0.826         & 0.816                           & 0.754    & 0.710 & 0.621 & 0.689 & 0.608 & 0.621 & 0.727 & 0.689 & 0.699  \\
          & $C_{l}AE-RF$ & 0.769 & 0.777         & 0.769                           & 0.752    & 0.704 & 0.608 & 0.620 & 0.477 & 0.507 & 0.661 & 0.610 & 0.623  \\\hline\hline
10        & $C_{o}AE-PC$ & 0.880 & 0.886         & \textbf{0.878} & 0.815    & 0.805 & 0.722 & 0.848 & 0.765 & 0.787 & 0.854 & 0.794 & 0.811  \\
          & $C_{l}AE-LR$ & 0.867 & 0.875         & 0.865                           & 0.777    & 0.804 & 0.721 & 0.779 & 0.689 & 0.717 & 0.822 & 0.814 & 0.818  \\
          & $C_{l}AE-RF$ & 0.828 & 0.842         & 0.830                           & 0.783    & 0.806 & 0.730 & 0.757 & 0.632 & 0.679 & 0.796 & 0.793 & 0.795 \\ \hline
\end{tabular}
\label{tab:fs}
\textbf{Legend:} FS -- Feature Selection; $W_P$ -- Weighted Precision; $W_R$ -- Weighted Recall; $W_F$ -- Weighted F-Measure.  
\end{table*}}

For all of the criteria (except criteria 6 and 7), we observed that, Random Forest classifier always misclassified the minority class (e.g., for criterion 1, minority class was `Satisfactory') into majority class (e.g., for criterion 1, majority class was `Not Satisfactory') which results a drop in recall value for the minority class. This happened due to the imbalanced class distribution of our dataset (see Fig. \ref{criteria_distribution}). In our dataset, criteria 6 and 7 are very close to balance and thus Random Forest classifier performed moderately.

\subsubsection{Identify Feature Size}
We have also varied the feature sizes to see how this impact on the classification performance.
In this part of the experiment, we have used the Pearson's Correlation feature selection method ($C_oAE-PC$) with the SVM classifer, as we found them best for our classification problem (see Table \ref{tab:fs}). Figure \ref{fig:fs} shows the performance of the SVM classifier combined with $C_oAE-PC$ feature set under various feature sizes -- 1000, 2000, 3000, 4000, 5000, 10000, and 53012.

\begin{figure*}[hbt!]
 \centering
     \includegraphics[width=\textwidth]{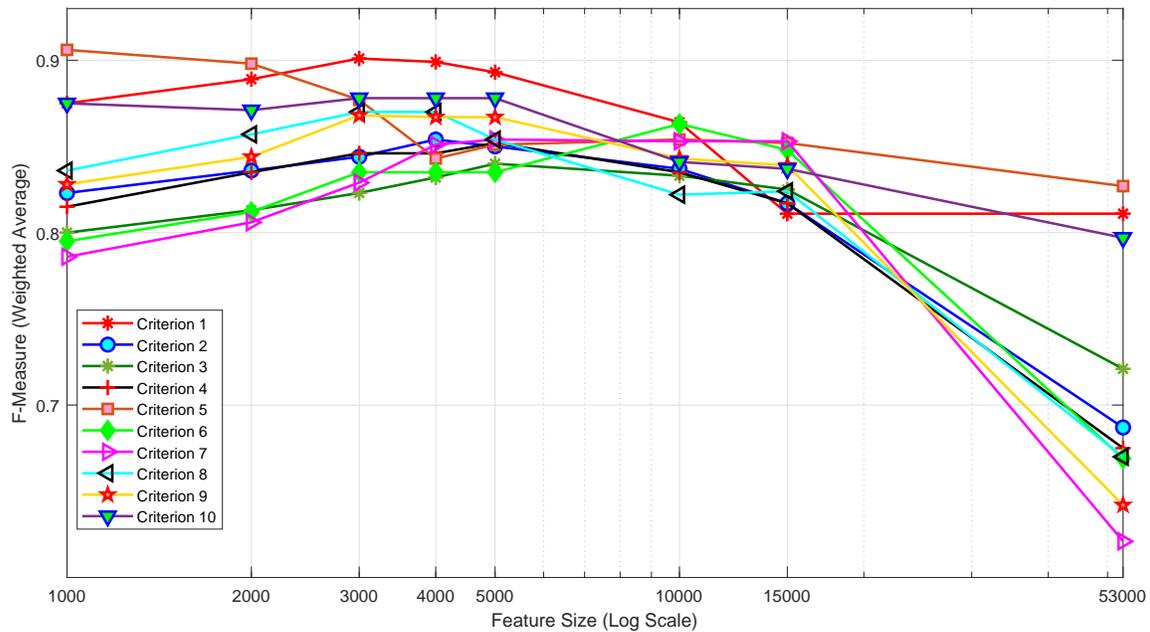}
    \caption{Changes of performance with respect to feature Size (Criteria 1-10)}
    \label{fig:fs}
\end{figure*}

We observed from the Figure \ref{fig:fs} that, for all criteria, the feature set comprises all features (total $53012$) performs lower due to having irrelevant and redundant features. We can see that there is an improvement in performance with reduced feature subset. For criterion 1, we achieved $90\%$ accuracy for the feature size $3000$ in terms of F-measure. For criteria $2, 4$ and $7$ , we achieved $85\%$ accuracy for the feature size $5000$. For criteria 3, we achieved $84\%$ accuracy for the feature size $5000$. Criterion 6 achieved $86\%$ accuracy for $1000$ sized feature set. For criterion 8 and 9, $87\%$ and $86\%$ accuracy were achieved for the feature size $4000$ and $3000$ respectively. For the rest two criteria ($5$ and $10$), we noticed that the performance curves are a bit different from other curves because of their imbalanced dataset nature. Criterion $5$ begins with the highest accuracy ($90\%$) at feature size $1000$ and performance varied with feature size. Criterion $10$ achieved highest accuracy ($88\%$) for the feature size $5000$. Overall, all reduced features subset (varied in size) achieved at least $80\%$ accuracy with our explored feature combination. 

\subsection{Class Balancing}
We noticed some imbalanced class in our dataset.
As we found that criteria $5$ and $10$ are most imbalanced class distribution, we have combat this class imbalance problem by adopting three class balancing techniques – Under sampling, Over-sampling and Synthetic Minority Over-Sampling Technique (SMOTE). As over-sampling duplicates the minority class instances, it can lead to model over-fitting. Similarly, under sampling can degrade performance if it leaves out important instances while cutting down. Thus, we also experimented our dataset with SMOTE which generates synthetic sample of minority class rather than using duplicates. However, SMOTE still does not prevent over-fitting as it generates synthetic data from existing data points. 

\begin{figure*}
\begin{subfigure}[t]{0.5\textwidth}
\includegraphics[width=3.3in,height = 2.1in]{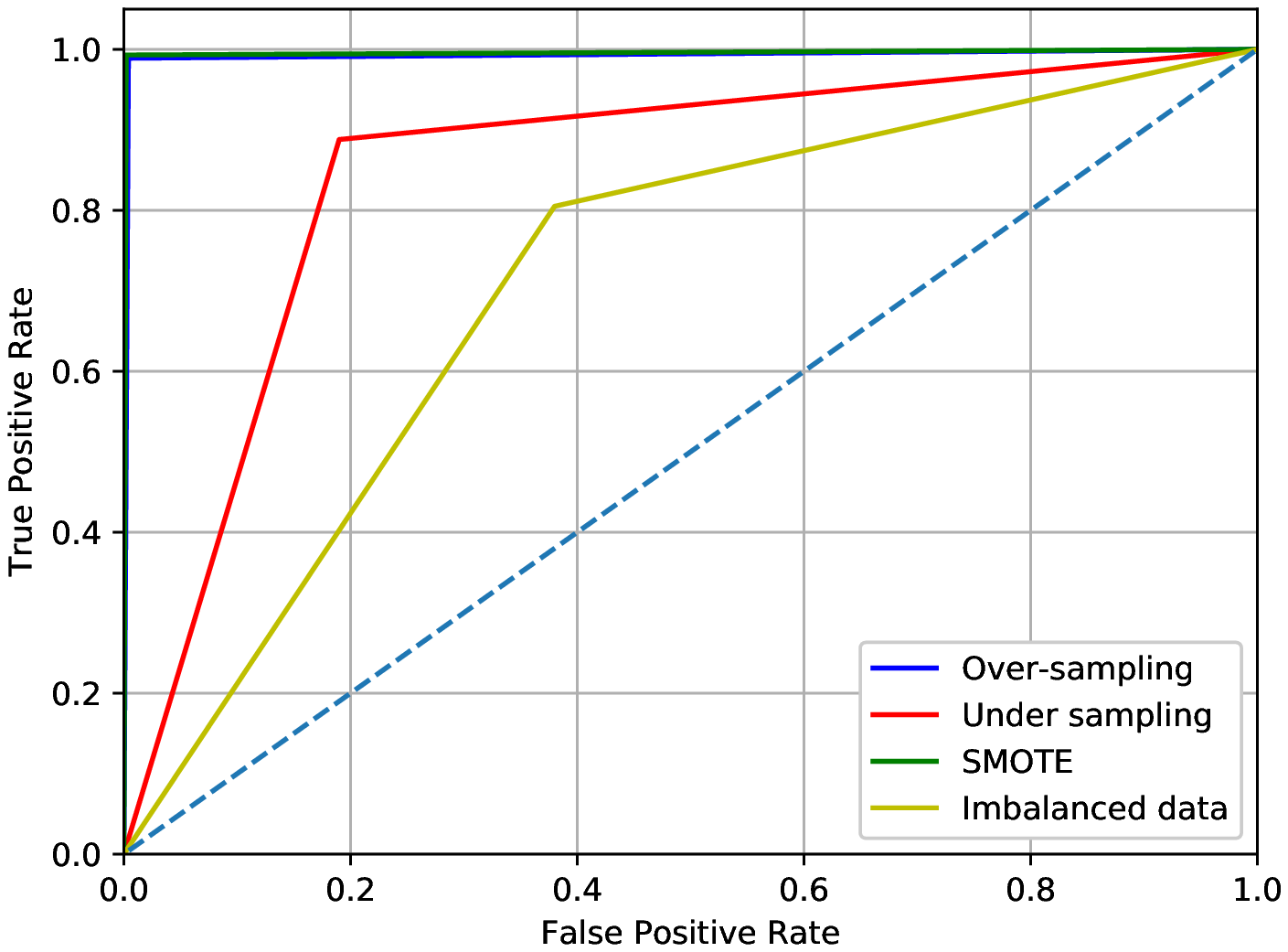}
\caption{Criterion 5}
 \label{fig:roccr5}
\end{subfigure}
\hspace{\fill}
\begin{subfigure}[t]{0.5\textwidth}
\includegraphics[width=3.3in,height = 2.1in]{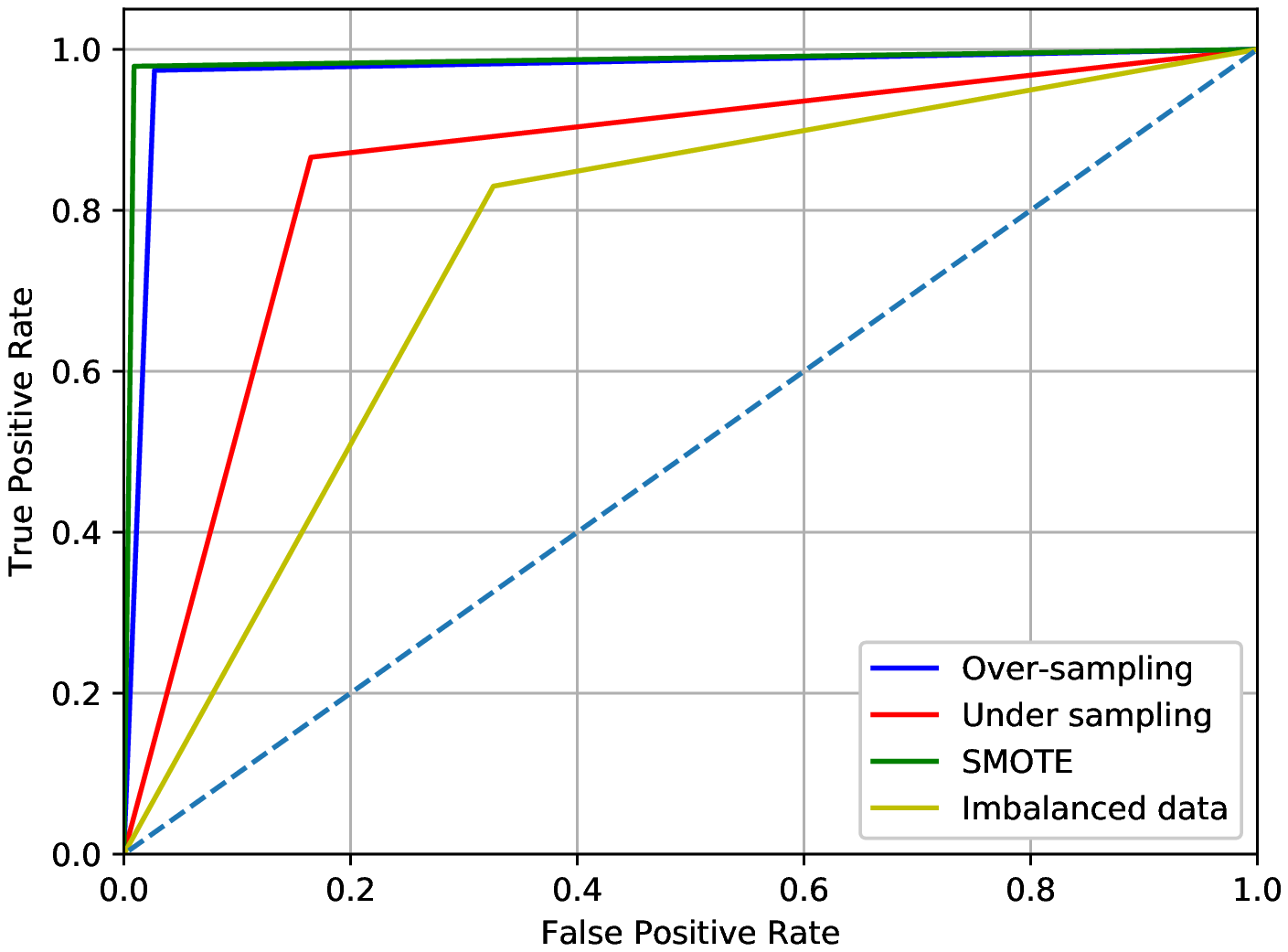}
\caption{Criterion 10}
 \label{fig:roccr10}
\end{subfigure}
\caption{ The ROC curve of balanced and imbalanced class for the feature size $5000$ (Criteria 5 and 10)}
\label{fig:roc}
\end{figure*}

Figure \ref{fig:roc} shows the performance comparison of these three methods in terms of Receiver Operating Characteristics (ROC) curve. It is observed that, all of the sampling techniques performed better than the imbalanced dataset. For SMOTE, we have got $99\%$ and $98\%$ accuracy for criteria 5 \& 10, respectively. 

\section{Semantic Analysis of Features}
\label{sec-agreement}
In this section, we have semantically analyzed the co-relations between a criterion and its corresponding most significant feature set to show how the feature set justify our assessment of a criterion. 

To analyze each criterion, we have itemized the top 16 most discriminating features by combining the results found from Pearson's correlation, Logistic Regression and Random Forest feature selection algorithms. The top $16$ feature list is presented in the Table \ref{tab:topf}. Overall, POSWord count and TF-IDF features are found most significant and other features get varied from criteria to criteria. Insights gained from determining relevant features are as follows:
{\renewcommand{\arraystretch}{.2}
\begin{table}[]
\caption{Most discriminating Features [Criteria 1-10]}
\label{tab:topf}
\begin{tabular}{|m{.4cm}|p{7.6cm}|}\hline
Cr & Most Correlated Features (Top 16) \\ \hline\hline
1             & $Money^\star$, $\mathbf{cost\_NN^\ddagger}$, $\mathbf{costs\_VBZ ^         
                \ddagger}$,$\mathbf{insurance\_NN^\ddagger}$, $\mathbf{costs\_NNS^\ddagger}$, $But\_CC^\ddagger$,      $\mathbf{price\_NN^\ddagger}$, $negate^\star$,$ verb^\star$, $ not\_RB^\ddagger$,  $\mathbf{ dollars\_NNS^\ddagger}$, $\mathbf{covered\_VBN^\ddagger}$,                     $\mathbf{thousand\_dollar^\ast}$, $\mathbf{ pay\_VB^\ddagger},$      $\mathbf{cost\_VB^\ddagger} $, $tag\_NN^\ddagger$\\\hline
2         &     $\mathbf{Percent\_NN^\ddagger}$, $quant^\star$ ,$were\_VBD^\ddagger$,      
                $compared\_VBN^\ddagger$,$\mathbf{england\_journal^\ast}$, $\mathbf{group\_NN^\ddagger}$, $differ^\star$ ,$new\_england^\ast$, $\mathbf{year\_percent^\ast}$, $\mathbf{Reuters\_NNP^\ddagger}$, $standardsth\_thomson^\ast$, $trust\_principl^\ast$, $\mathbf{compar\_percent^\ast}$, $reuter\_trust^\ast$, $\mathbf{percent\_percent^\ast}$, $journal\_medicin^\ast$                     \\\hline
3         & $\mathbf{ not\_RB^\ddagger}$,
            $ \mathbf{should\_MD^\ddagger}$,
            $WC^\star$,
            $negate^\star $,
            $differ ^\star$,
            $ \mathbf{effects\_NNS^\ddagger}$,
            $ some\_DT^\ddagger$,
            $ cause_VB^\ddagger$,
            $ \mathbf{risks\_NNS^\ddagger}$,
            $tentat ^\star$,
            $ \mathbf{have\_VB ^\ddagger}$,
            $ \mathbf{nausea\_NN^\ddagger}$,
            $\mathbf{external\_link^\otimes}$,
            $ common\_effect^\ast$,
            $ effect\_includ^\ast$,
            $ side\_JJ\ddagger$,
            $high\_dos^\ast$,
            $ died\_VBD^\ddagger$
                      \\\hline
4         &    $\mathbf{ study\_NN^\ddagger}$,
                $Normalizeddistinctwordcount^\otimes$,
                $ \mathbf{not\_RB^\ddagger}$,
                $\mathbf{ randomly\_RB^\ddagger}$,
                $differ^\star$ ,
                $\mathbf{assigned\_VBN^\ddagger}$,
                $\mathbf{studies\_NNS^\ddagger}$,
                $The\_DT^\ddagger$,
                $\mathbf{were\_VBD^\ddagger}$,
                $\mathbf{placebo\_NN^\ddagger}$,
                $ One\_CD^\ddagger$,
                $\mathbf{editorial\_NN^\ddagger}$,
                $\mathbf{evidence\_NN^\ddagger}$,
                $group\_NN^\ddagger$,
                $randomized\_VBN^\ddagger$,
                $random\_assign^\ast$,
                $placebo\_group^\ast$,
                $ email\_NN^\ddagger$
                  \\\hline
5         & $famili\_histori^\ast$,
            $\mathbf{Anesthesiologists\_NNPS^\ddagger}$,
            $\mathbf{dri\_eye^\ast}$,
            $revealed\_VBN^\ddagger$,
            $history\_NN^\ddagger$,
            $\mathbf{Hed\_NNP^\ddagger}$,
            $\mathbf{moist\_JJ^\ddagger}$,
            $\mathbf{transit\_NN^\ddagger}$,
            $american\_suffer^\ast$,
            $\mathbf{anesthesiology\_NN^\ddagger}$,
            $need\_new^\ast$,
            $excessive\_JJ^\ddagger$,
            $inform\_patient^\ast$,
            $\mathbf{labbased\_JJ^\ddagger}$,
            $histori\_breast^\ast$,
            $air\_NN^\ddagger$
                      \\\hline
6         & $\mathbf{per\_ner\_count^\otimes}$,
            $\mathbf{professor\_NN^\ddagger}$,
            $\mathbf{study\_NN^\ddagger}$,
            $University\_NNP^\ddagger$,
            $\mathbf{involved\_VBN^\ddagger}$,
            $\mathbf{The\_DT^\ddagger}$,
            $normalizeddistinctwordcount^\otimes$,
            $WC^\star$,
            $\mathbf{But\_CC^\ddagger}$,
            $that\_IN^\ddagger$,
            $\mathbf{said\_VBD^\ddagger}$,
            $\mathbf{National\_NNP^\ddagger}$,
            $School\_NNP^\ddagger$,
            $\mathbf{not\_RB^\ddagger}$,
            $funded\_VBN^\ddagger$,
            $about\_IN^\ddagger$
                      \\\hline
7         & $\mathbf{Not\_RB^\ddagger}$,
            $\mathbf{But\_CC^\ddagger}$,
            $WC^\star$,
            $Differ^\star $,
            $\mathbf{There\_EX^\ddagger}$,
            $\mathbf{The\_DT^\ddagger}$,
            $\mathbf{Than\_IN^\ddagger}$,
            $\mathbf{Are\_VBP^\ddagger}$,
            $normalizeddistinctwordcount^\otimes$,
            $\mathbf{Many\_JJ^\ddagger}$,
            $For\_IN^\ddagger$,
            $\mathbf{Year\_NN^\ddagger}$,
            $That\_DT^\ddagger$,
            $University\_NNP^\ddagger$,
            $Often\_RB^\ddagger$,
            $Better\_JJR^\ddagger$
                      \\\hline
8         & $Not\_RB^\ddagger$,
            $differ^\star $,
            $are\_VBP^\ddagger$,
            $negate^\star$,
            $radiotherapy\_NN^\ddagger$,
            $Alessandro\_NNP^\ddagger$,
            $\mathbf{Magnet\_reson^\ast}$,
            $\mathbf{CITATION\_NNP^\ddagger}$,
            $\mathbf{Twice\_week^\ast}$,
            $\mathbf{Reson\_imag^\ast}$,
            $Temperature\_NN^\ddagger$,
            $Outcom\_studi^\ast$,
            $Resonance\_NN^\ddagger$,
            $Cognit\_impair^\ast$,
            $\mathbf{Welltolerated\_VBN^\ddagger}$,
            $\mathbf{Axis\_NN^\ddagger}$
                  \\\hline
9         & $Have\_VBP^\ddagger$,
            $Studies\_NNS^\ddagger$,
            $\mathbf{FoxNewscom\_NNP^\ddagger}$,
            $\mathbf{Moisturizers\_NNS^\ddagger}$,
            $News\_releas^\ast$,
            $Result\_promis^\ast$,
            $Help\_woman^\ast$,
            $\mathbf{Control\_blood^\ast}$,
            $Consumption\_NN^\ddagger$,
            $\mathbf{Healing\_VBG^\ddagger}$,
            $\mathbf{Leadership\_NN^\ddagger}$,
            $Molecular\_JJ^\ddagger$,
            $\mathbf{Melbourne\_NNP^\ddagger}$,
            $\mathbf{Educated\_VBN^\ddagger}$,
            $Obesity\_NNP^\ddagger$,
            $\mathbf{Penetrate\_VB^\ddagger}$
                      \\\hline
10        & $Differ^\star$,
            $Negate^\star$,
            $social ^\star$,
            $tentat^\star$,
            $Sixltr^\star$,
            $\mathbf{Detect\_diseas^\ast}$,
            $News\_releas^\ast$,
            $Develop\_research^\ast$,
            $Lead\_investig^\ast$,
            $\mathbf{Collaborate\_VBP^\ddagger}$,
            $\mathbf{Discovery\_NN^\ddagger}$,
            $\mathbf{Tumour\_NN^\ddagger}$,
            $Media\_contact^\ast$,
            $Innovator\_NN^\ddagger$,
            $\mathbf{Resume\_VB^\ddagger}$,
            $\mathbf{Exceptional\_JJ^\ddagger}$
                       \\ \hline
 
\end{tabular}
\newline\\
\textbf{Legend:} Cr -- Criterion $\star$ -- LIWC Feature; $\ast$ -- TF-IDF Feature; $\ddagger$ -- POSWord count; $\otimes$ -- Miscellaneous Features; Features common in all three feature set are indicated by Bold texts. 

\end{table}}
 
As criterion 1 is about coverage of cost intervention, it is instinctive to have features associated with money, cost, price, dollars, amount of dollars (thousand, hundred), insurance etc.; each of which are found as top discriminating features in this study.

Inclusion of absolute number in quantifying benefit gives readers a better sense of understanding about an intervention. For example, the sentence `New drug reduces heart failure risk in half' can give reader a peachy idea about the intervention. But if the sentence was like `$4\%$ risk dropping to a $2\%$' (showing risk halved), it would sound less significant to the reader with clear idea. From the top selected feature subset for criterion 2, we find TF-IDF feature `percent' and LIWC feature `number' to be pertinent to the usage of absolute number in article. Besides, TF-IDF feature `compar', `trust'; LIWC feature `differ', `quant' are also relevant in explaining benefits. 

When reading a story about a new intervention, it is expected to have explanation about the potential harms and side effects of the intervention. In our feature set, we have found POSWord feature `risks', `cause', `nausea' (a common side effect of drug), `side', `died'; TF-IDF feature `common', `effect', `high'; and LIWC feature `negate', `tentat' – are quite meaningful to describe criterion 3.

In order to grasp the quality of the evidence, a story needs to present an elaborate explanation of the study (source, size, type, limitation, etc.) it went through. For example, a report published in The Wall Street Journal on Ebola Vaccine stated that this was the `first placebo-controlled study of two vaccines against the Ebola virus' and mentioned its shortcomings as well and the reviewers in the HealthNewsReview.org rated it as `Satisfactory' for criterion 4. Our feature subset consists of POSWord Features `randomly', `assigned', `placebo', `study', `evidence' and `group', and TF-IDF feature `random-assign' are stringently aligned with this criterion in describing evidentiary details. 

It is a matter of judgement to identify disease mongering. From the feature subset found for criterion 5, we can relate TF-IDF features `dry-eye', `suffer', `need', `inform', and POSWord features `revealed', `excessive' are directly inflating the seriousness of a condition. For example, using rating scales to diagnose chronic dry-eye is simply an exaggeration of a common disorder. As only $19\%$ articles from our dataset were rated `Not Satisfactory' on this criterion, we found less aligned extracted features to define disease mongering.

According to criterion 6, independent experts should be included in news stories about health care interventions and conflicts of interest in the people who are quoted should be explored and disclosed. In order to explore this criterion, we defined a new feature, `per\_ner\_count' to count the number of person referred in a document and we found this feature to be the most relevant feature to describe this criterion. Same is the case with the feature `Org\_ner\_count' which gives us the count of organizations cited in a document. Apart from these, POSWord features – `university', `that', `said', `national', `professor', `study', and `involve' are also aligned with this criterion to describe it.

As criterion 7 is about comparing new intervention with existing alternatives, it is usual for a document to contain comparison words. From our feature subset, we found that the LIWC feature  `differ', and POSWord feature `than', `not', and `but' are more relevant to describe this criterion.

We observe that the feature subsets we found for the criteria 8, 9 and 10 could not properly describe the properties of these criteria. We found that only $25\%$, $23\%$ and $8\%$ stories from the criteria 8, 9 and 10 respectively were rated with ` Not satisfactory’ in our dataset which may have brought up the reason for our feature selection algorithm to fail in differentiating the discriminating features .

\section{Discussion}
\label{sec-discussion}

In this study, we have examined the application of machine learning approach to automate the quality assessment process for web based health related information. We found that it is feasible to apply machine learning classifiers to estimate the quality of health related articles if the classifier can be trained properly. This work is not directly comparable to the already existing studies because most of the studies examined the quality of health information from a single domain perspective (e.g., vaccination \cite{l43}, \cite{l44}, \cite{l45}; diabetic neuropathy \cite{l11}; reproductive health information \cite{l41}; nutrition coverage \cite{l42} etc.) through a manual process and statistical analysis. We have examined articles over entire health domain, ensuring its applicability to all possible health related category. In this context, our work will make manual reviewing process scalable and save manual labour and time. Our developed dataset will help researchers to contribute in the growing field of health care research. Overall, this automated quality assessment approach may help search engine to promote high quality health information and discourage low quality articles. 

However, there are some limitations in our study. Experts from HealthNewsReview.org used three labels - `Satisfactory', `Not Satisfactory' and `Not Applicable' for characterizing $10$ criteria . Cases where a number of criteria may be impossible or unreasonable for some of the stories were rated as `Not Applicable' by the review experts. In our study, we deducted stories with `Not Applicable' criteria from our training set as those stories constituted a small part of the whole corpus and trained our classifiers for two class labels - `Satisfactory' and `Not Satisfactory'. That's why we could not use all $1720$ articles for each of the $10$ criteria and number of total dataset varied from criteria to criteria (e.g., our dataset for criterion 1 comprised of $1426$ articles after removing class instances of `Not Applicable' label).  In our future study we plan to address this shortcoming.

Another limitation is, our dataset is not large enough to be compatible for deep learning framework. We trained deep learning classifier for our dataset though and found approximately $50\%$ accuracy over all criteria. In our future work, we plan to enrich our dataset to examine its feasibility from deep learning perspective.

\section{Conclusion and Future Work}
\label{sec-conc}
In this paper, we have applied data mining approach to automatically assess the quality of online health articles. We have prepared our dataset comprises $1720$ health related articles extensively reviewed by a group of experts. Through a pipeline of data pre-processing steps, we have refined our data and extracted $53012$ features to train classifiers. We have identified the best feature selection technique to select most relevant feature subset from our feature space, and have applied four different classifiers - SVM, Naive Bayes, Random Forest and EnsembleVote to train model. For our dataset, we found SVM is the best performer achieving accuracy upto $84\%$ to $90\%$ for ten different criteria. We have also analyzed top 16 most correlated features for each of the ten criteria to justify the feasibility of our assessment. We found that our selected features are capable of characterizing criteria successfully.
From our experimental results and analysis, it can be concluded that it is feasible to apply data mining techniques to automate quality assessment process for online health articles. Following the richness of dataset and specific focus independent nature of analysis, proposed model may serve as a universal standard for appraising quality of OHA and wipe out the negative impact of misinformation dissemination to some extent. 

As future work, we will further investigate this study with deep learning approach. We have also plan to explore multinomial classification problem to evaluate health related articles which cannot address some of the specific criteria.

\ifCLASSOPTIONcaptionsoff
  \newpage
\fi



%

\ifCLASSOPTIONcaptionsoff
  \newpage
\fi

\bibliographystyle{IEEEtran}\bibliography{main}

\vfill

\end{document}